\documentclass[aps,prb,twocolumn,showpacs,superscriptaddress]{revtex4}
\usepackage{bm,amsmath,amssymb,latexsym,mathrsfs,enumerate}
\usepackage[mathcal]{euscript}
\usepackage{graphicx, subfigure}
\newcommand{\figwidth}{0.47\textwidth}

\begin{document}

\title{Vortex ground state for small arrays of magnetic
particles with dipole coupling}

\author{S.~A. Dzian}
\affiliation{National Taras Shevchenko University of Kiev, 03127
Kiev, Ukraine}

\author{A.~Yu. Galkin}
\affiliation{Institute of Metal Physics, 03142 Kiev, Ukraine }

\author{B.~A. Ivanov}
\email{bivanov@i.com.ua} \affiliation{Institute of Magnetism, 03142,
Kiev, Ukraine} \affiliation{National Taras Shevchenko University of
Kiev, 03127 Kiev, Ukraine}

\author{V.~E. Kireev}
\affiliation{Institute of Magnetism, 03142, Kiev, Ukraine}

\author{V.~M. Muravyov}
\affiliation{Institute of Magnetism, 03142, Kiev, Ukraine}

\begin{abstract}
We show that a magnetic vortex is the ground state of an array of magnetic particles
shaped as a hexagonal fragment of a triangular lattice, even for an
small number of particles in the array $N \leq 100$. The vortex core
appears and the symmetry of the vortex state changes with the
increase of the intrinsic magnetic anisotropy of the particle $\beta$;
the further increase of  $\beta$ leads to the destruction of the
vortex state. Such vortices can be present in arrays as small in size
as dozen of nanometers.
\end{abstract}
\pacs{75.10.Hk, 75.50.Tt, 75.30.Kz}


\maketitle

\section{Introduction}
\label{sec:intro}

Topological defects of vortex type play a paramount part in the
general physics of ordered media such as superfluidity,
superconductivity and magnetism. In particular, vortices and vortex
pairs are important in two-dimensional (2D) magnetism. Recently, the
ground state of  soft ferromagnetic particles of micron and
sub-micron size  has been shown to be of a vortex type, which  has
received much attention from the research
community.~\cite{Skomski,Antos+Rev,GusObzor} Compared to vortices in
superfluid systems, magnetic vortices in 2D ferromagnets and
antiferromagnets have a richer behavior, since they may be divided
into two different classes, in-plane and out-of-plane
vortices.~\cite{IvMert98,Mert+Obzor} For in-plane vortices all spins
lie in the vortex plane. Out-of-plane vortices have nonzero spin
components orthogonal to the vortex plane localized within the
so-called \textit{vortex core}, a small region near the vortex
center. Vortices with a core are described by several different
types of topological
charges.~\cite{VolovMin77,MerminTopol,KosIvKovPhRep}  Besides the
standard $\pi_1$-topological charge~\textit{vorticity} $q$, which is
similar to \textit{circulation} in a supercurrent systems, one can
introduce $\pi_2$-topological charge, the \textit{polarity} or
\textit{polarization} $p=\pm 1$ of the vortex core, which is the
spin direction in the core and is connected to the $\pi_2$
topological charge of the magnetization field. In-plane vortices can
be associated with the value $p=0$. Further, vortices found in the
ground state of soft magnetic particles should have $q=1$ only
($q=-1$ corresponds to antivortices that can be connected to
``antidots'', small holes in a patterned magnetic
film~\cite{HertelAntiVort}),  but they are additionally classified
by the discrete number \textit{chirality} $C=\pm 1$, which is the
sense of rotation of magnetization far from the vortex core. For a
single vortex in a bulk 2D magnet with easy-plane magnetic
anisotropy, there is a transition from coreless in-plane vortex
structure to the vortex with a well defined core, as the anisotropy
strength decreases below a certain critical
value.~\cite{WysinCritAniz} The presence of a core plays a crucial
role in the dynamic properties of magnetic vortices. In particular,
the value of $\pi _2$-topological charge determines special
gyroscopic properties of the vortex dynamics and the presence of
low-frequency dynamics, essential for possible applications of
magnetic vortices in perspective
spintronic\cite{Krivorotov_vort07,IvZaspPRL07,Khvalkov_PRB09,Iv+zvezdPZh10,Kim_vortSPC_PRB09,Mistral_generator,Cherepov+PRL12}
and magnonic~\cite{Magnonics,magnonics} devices.

The presence of vortices in the ground state of soft ferromagnetic
particles is determined by the balance between magnetostatic and
exchange energies. For disk-shaped particles with negligible
magnetocrystalline anisotropy and typical thicknesses about
$20\div50$~nm, the vortex state is stable if the disk radius exceeds
some critical value $R_c \sim 150\div200$~nm.  Vortices in the
ground state of soft magnetic particles possess a core with the size
of the order of the exchange length of the material (about $15$~nm
in Permalloy). It has been recently shown that magnetic vortices of
different structure can be the ground state for magnetic particles
with comparable energies of the exchange and dipolar interactions,
even for small (of the order of $10^{3}$) number of magnetic moments
in the particle.~\cite{IvKireevPZh11} Highly non-uniform ground
state is found for small particles with a sufficiently high surface
anisotropy as well.~\cite{DimWysin94+95,IvKireevPRB03}

From the perspective of a search for vortex states, the promising
magnetic systems are those in which the exchange interaction
suppressed or non-existent and the main source of interaction among
structural elements is the dipole-dipole interaction of their
magnetic moments. These are so-called \textit{dipolar magnets}, i.e.
such spin systems where a long-range magnetic dipole interaction
prevails. Dipolar magnets have been attracting persistent interest
during decades as objects of the fundamental physics of magnetism
possessing some unusual properties. One may mention the presence of
an ambiguous ground state with non-trivial continuous degeneracy
even for simple cubic\cite{LutTisza,BelobGIgnat} or 2D square
lattices,\cite{DipD2,Gus99,dipObzorUFN} and the existence of special
phase transitions induced by an external magnetic
field.~\cite{BishGalkIv,GalkIvPZh06} Magnon spectra of such systems
have a non-analytic behavior at  small wave vectors.
\cite{Karet+Fraerman,PolitiPini02,GiovanniniBloch,Galkin+Vort,Tartak+08,Bondarenko+prb12}

The interest in systems with the dominant magnetic dipole
interaction has significantly increased in last years, mainly in the
context of artificial magnetic materials such as arrays of magnetic
nanoparticles.~\cite{Skomski} Magnetic systems with the dominant
dipole interaction possess interesting physical properties important
for applications. Among those properties, one can highlight the fact
that the ground state of an infinite system of magnetic moments
constituting a lattice and coupled by the dipole-dipole interaction
depends essentially on the lattice structure,~\cite{dipObzorUFN} and
in the presence of the intrinsic (intra-particle) anisotropy it
depends on the orientation of the easy axis of this anisotropy  with
respect to the lattice axes as well.

Let us discuss 2D lattices, which will be the object of our study.
For the lattices of particles with a high perpendicular anisotropy,
the ground state corresponds to various types of two-sublattice
antiferromagnetic order, particularly, a chessboard structure is
realized for a square lattice,~\cite{BishGalkIv} and a layered one
for a triangular lattice.~\cite{KirIvanovPZh09} For finite fragments
of such lattices these structures vary insignificantly being
compared to the infinite case. In the case of systems with an
in-plane anisotropy, the role of the boundaries is more essential.
For an infinite square lattice, the ground state  has
four-sublattice antiferromagnetic order and has a high (continuous)
degeneracy, while for the triangular lattice a ferromagnetic order
is realized.~\cite{dipObzorUFN} The presence of a boundary, however,
may change significantly the state of such a system. For finite
fragments of the square lattice the aforementioned continuous
degeneracy is removed, but the state remains
antiferromagnetic.~\cite{IvLTP2005,GalkIvPZh06} For a triangular
lattice of dipoles, the state changes are more radical, and in
finite samples the ground state can be a vortex state with a closed
flux of magnetization formed by the magnetic moments lying in the
plane of the array.~\cite{PolitiPiniTreug} In general, the reason of
emergence of vortices is the same as for soft magnetic
dots~\cite{Usov}, see also,~\cite{Skomski,Antos+Rev,GusObzor} where
the vortex state emerges due to the energy gain of closure of
magnetic flux in the sample, and the vortices are an alternative to
a standard domain structure. However, for a rectangular-shaped array
the vortex state is advantageous only for large enough
arrays.~\cite{PolitiPiniTreug}

\begin{figure}
\includegraphics[width=\figwidth]{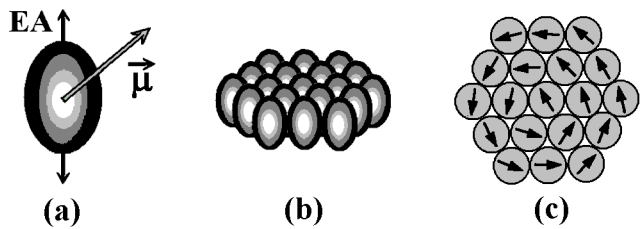}
\caption{\label{fig1}   The model of a planar array of magnetic
nanoparticles: (a) the shape of a single
elongated magnetic particle, the easy axis is denoted as EA,
orientation of the magnetic moment $\vec \mu$ is denoted by
the arrow; (b) a symmetric plane cluster comprised of 19
particles; (c) the same in a plane view; magnetic moments of
particles in a planar vortex existing at small anisotropy are
shown with arrows. }
\end{figure}

In view of the importance of vortices for the fundamental physics of
magnetism as well as for a variety of applications, search for
physical systems with vortex-type ground state is of great interest.
In particular, it is interesting what is the minimal size of a
system carrying a vortex, and whether  that size can be made
significantly smaller than the critical size $R_{c}$  indicated
above.

In this paper, we will demonstrate that for an array of particles
shaped as a  fragment of a triangular lattice with a high hexagonal
symmetry, see Fig.~\ref{fig1}, the ground state is a vortex state
even for a small number of particles in the array. Even for an
extremely small array of such type, which consists of 7 particles,
the vortex state energy is almost two times lower than that of a
quasi-homogeneous state. We show further, that such arrays have very
interesting behavior: their ground state structure is highly
sensitive to the change of  the anisotropy of a single particle. In
particular, for a particle made of a soft magnetic material this
anisotropy could be varied by changing the particle shape.

\section{Model and results}

Consider an array of particles placed at the sites of a finite
hexagonally shaped fragment of a 2D triangular lattice. The energy
of the this system contains contributions from the energy of the
magnetic dipole interaction and from the energy of the anisotropy:
\begin{multline}
\label{eq1} W=\sum\limits_{\vec l\ne \vec l'} {\frac{\vec \mu_{\vec
l} \vec \mu_{\vec l'} -3(\vec \mu_{\vec l}  \vec \nu )(\vec
\mu_{\vec
l'} \vec \nu )}{| {\vec l-\vec l'} |^3}} + \\
+\frac{\beta }{a^3}\sum\limits_{\vec l} {[(\vec \mu_{\vec l} \cdot
\vec e_x )^2+(\vec \mu_{\vec l} \cdot \vec e_y )^2]} ,
\end{multline}
where $\vec \mu_{\vec l} $ is the magnetic moment of the particle at
site $\vec{l}$, $\vert \vec \mu\vert =\mu _0 $, $\mu _0 $ is the
magnetic moment of a single particle, $\vec \nu =( {\vec l-\vec l'}
)/| {\vec l-\vec l'} |$, $a$ is the lattice constant (the distance
between closest particles in the array plane), and $\beta$ is a
dimensionless constant determining the magnetic anisotropy strength
of a particle. This anisotropy is assumed to be uniaxial, of the
easy-axis type (so that $\beta >0$), with the easy axis $\vec e_z $
perpendicular to the system plane. Here $\vec e_{x,y,z}$ are unit
vectors along coordinate axis.

One can expect that the absence or presence of a vortex core will depend
on the effective anisotropy of the system. It is important
to note that the total magnetic anisotropy of the array results from
the uniaxial anisotropy of separate particles and from the easy-plane anisotropy
of the array induced by the demagnetization field of a planar set of
magnetic moments (similar to the shape
anisotropy in the thin films). Competition of these two contributions determines a
complex character of the distribution of magnetization in the array.

If the easy-plane anisotropy induced by the demagnetization field is
sufficiently large, that leads to a suppression of the vortex core.
One can expect that the effective anisotropy can be changed by using
magnetic particles possessing their own intrinsic easy-axis
anisotropy. For the core to emerge, the effective anisotropy should
be reduced, which takes place for particles with the easy axis
perpendicular to the array plane. For arrays of particles made of
soft magnetic materials, such situation is realized in case of
elongated particles oriented perpendicular to the array plane, see
Fig.~\ref{fig1}. Such a geometry of the problem is promising for
ultradense information storage\cite{chou94,Meier98,Ross} and is
naturally realized, for example, if the array is created by
self-organization of small elongated particles floating in liquid.
Competition of these two magnetic interactions provides the
possibility to change the effective anisotropy of the system, which,
as we will show below, allows one to impact the structure of the
magnetic macrovortex.

\begin{figure}
 \subfigure[\ $ \beta =0$]{\includegraphics[width = 27mm]{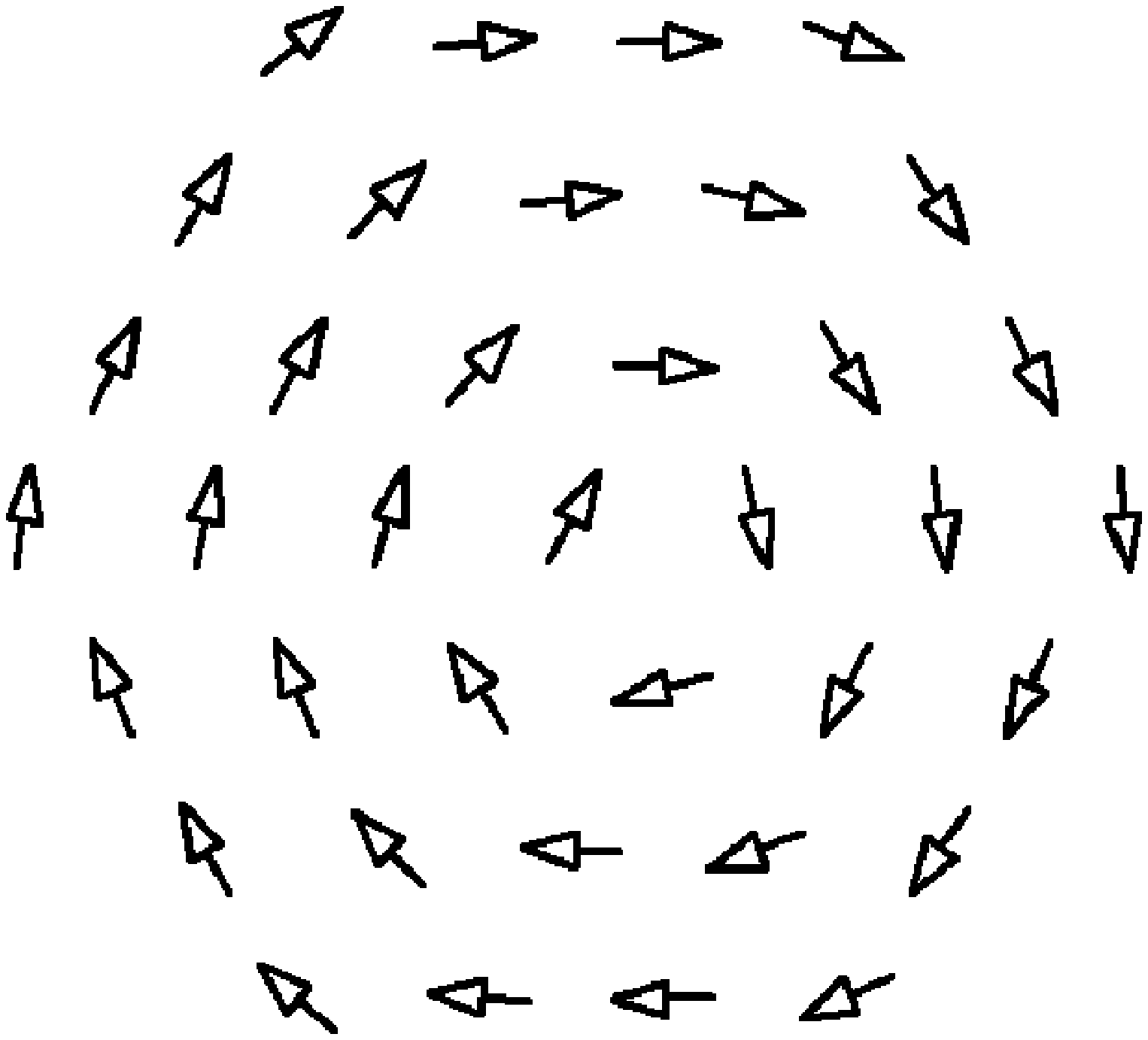}
 \label{1}}
 \subfigure[\ $ \beta=0.72$  ]{\includegraphics[width = 27mm]{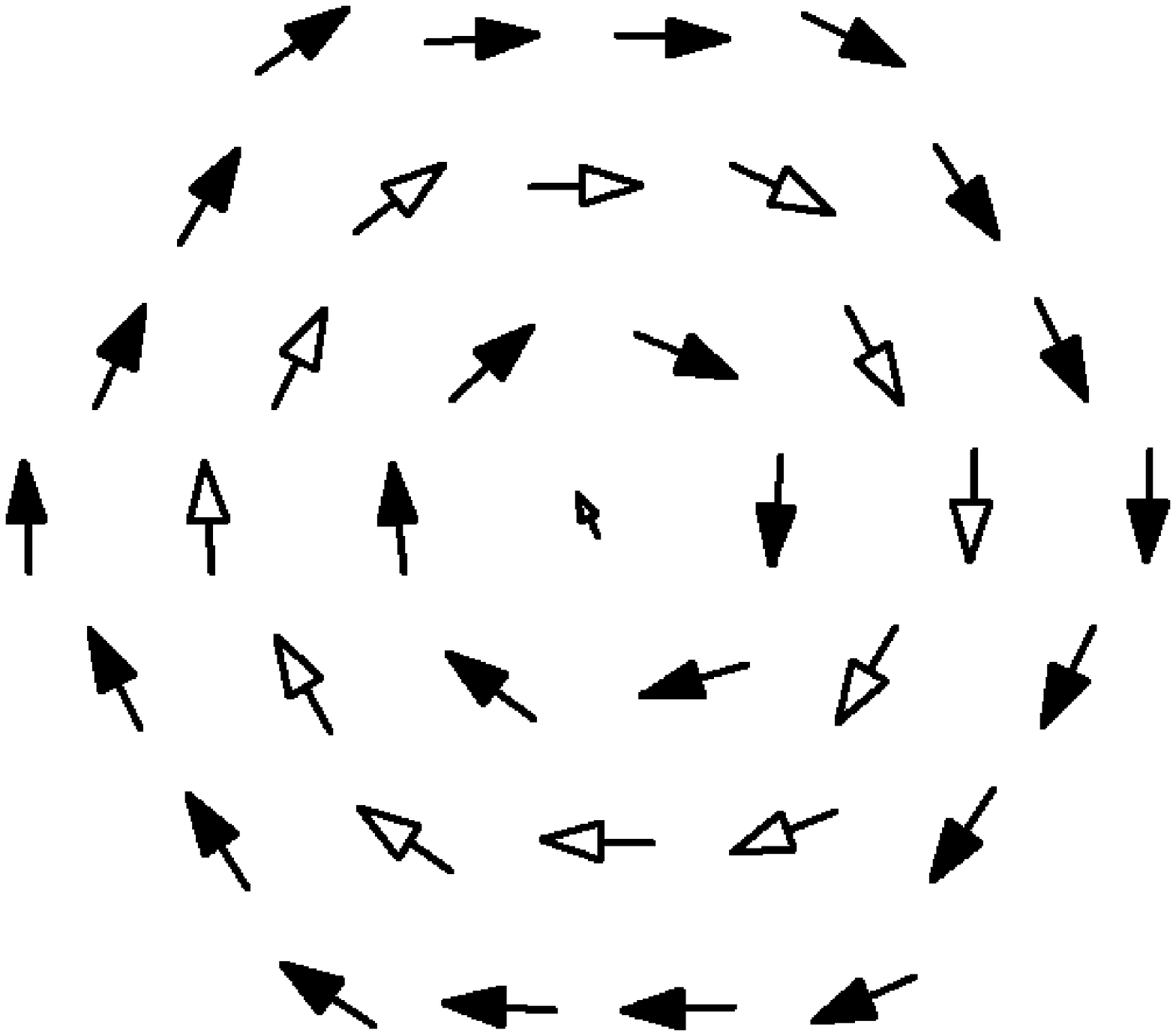}
 \label{2}}
 \subfigure[\ $ \beta =0.75$ ]{\includegraphics[width = 27mm]{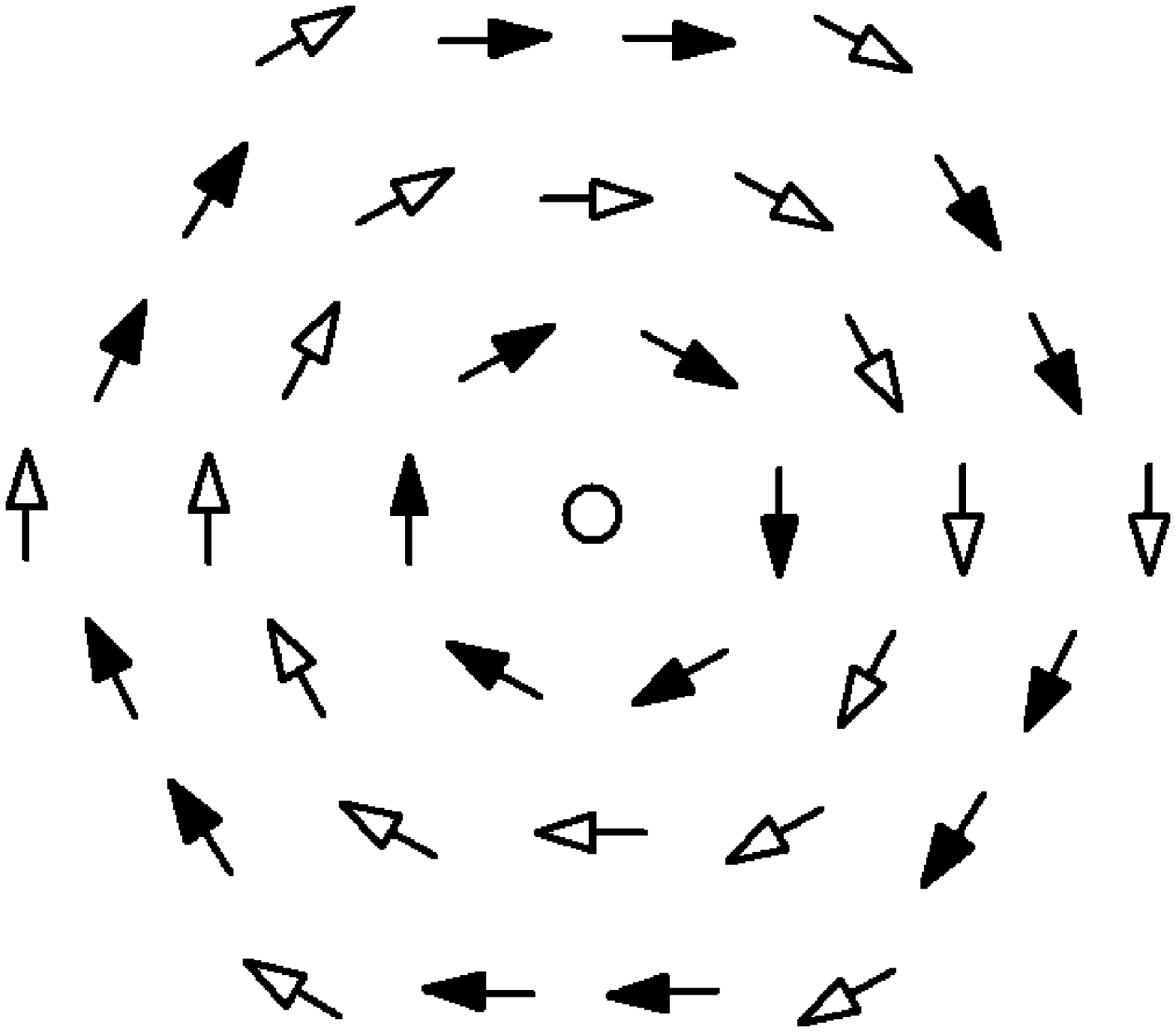}
 \label{3}}
 \subfigure[\ $ \beta =1.102 \simeq  \beta _t$ ]{\includegraphics[width = 27mm]{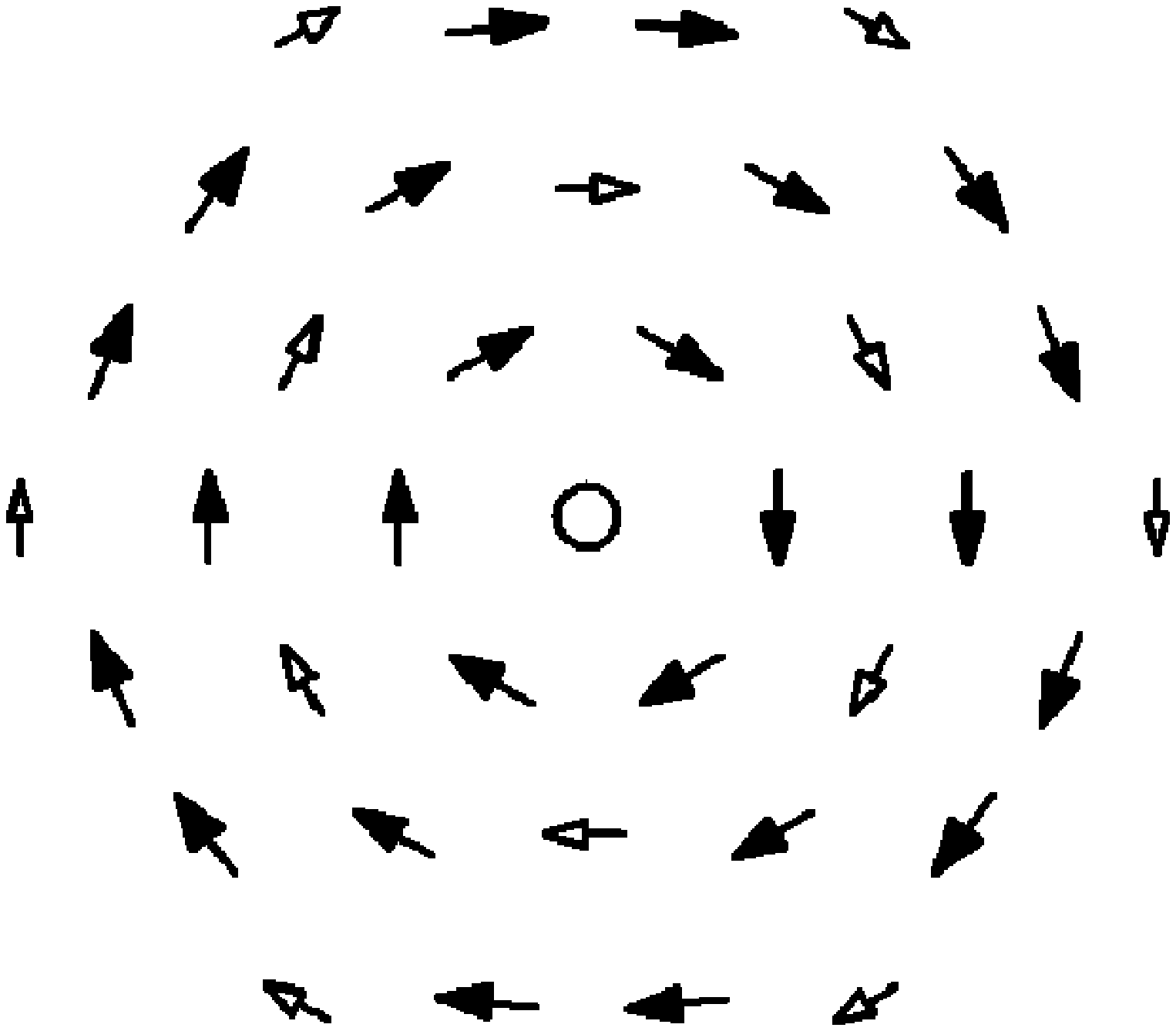}
 \label{4}}
 \subfigure[\ $ \beta =1.105$, antiferromagnetic state]{\includegraphics[width = 27mm]{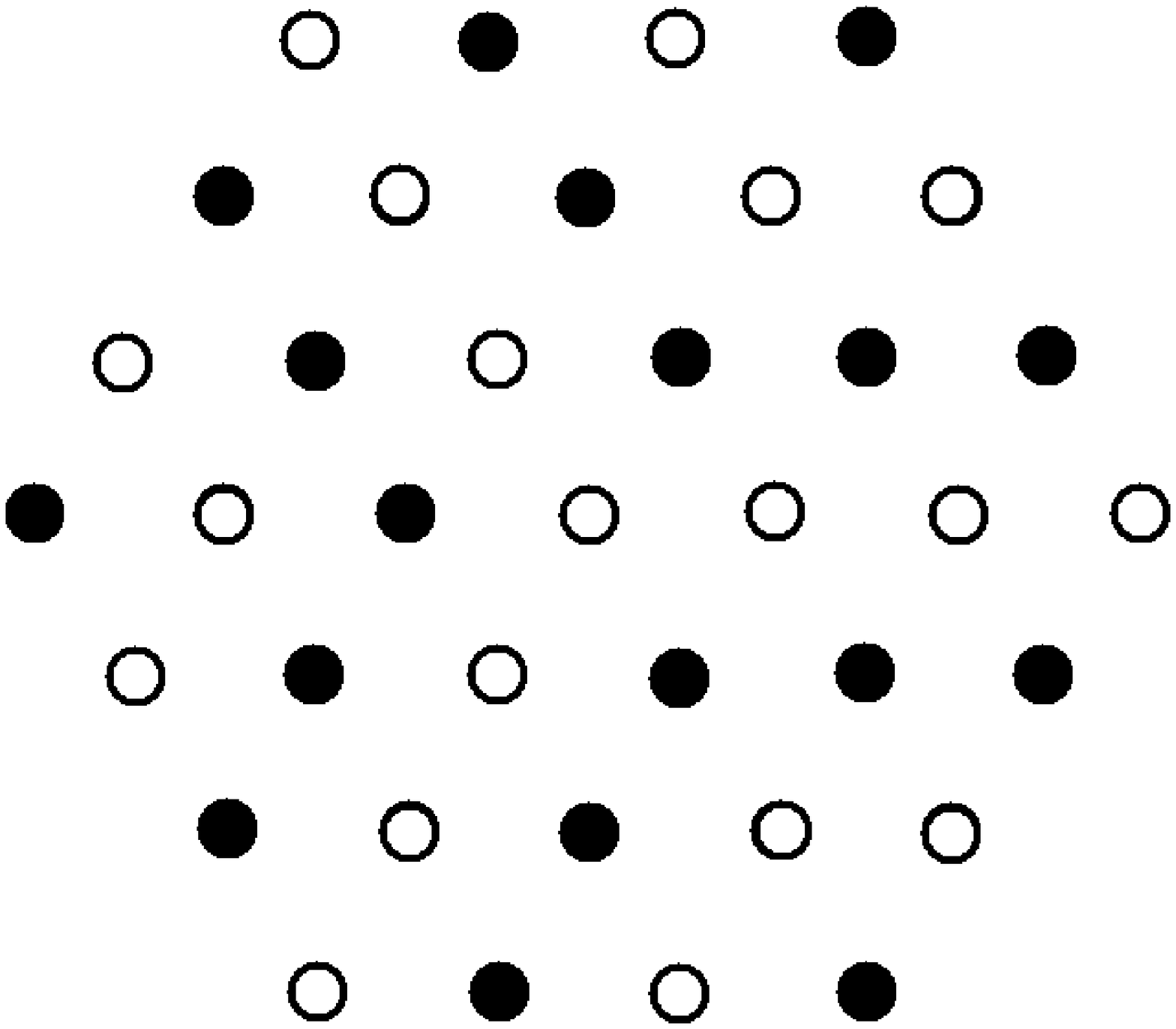}
 \label{5}}
\caption{Magnetic structure for the cluster of $N=37$ particles at
different values of anisotropy constant $\beta$, here $ \beta _1
=0.697$, $ \beta _2 =0.729$, see the text. The planar components of
the magnetic moments for each particle are shown with arrows, and
vertical moments  are represented by circles. For magnetic moments
with the positive or zero $z$-projection, open symbols (arrows with
open heads or open circles) are used, while particles with negative
$z$-projection of magnetic moments are depicted by the solid
symbols.} \label{fig2}
\end{figure}

To analyze this problem, we have employed two methods: numerical
minimization of the energy (\ref{eq1}) using the standard
Gauss-Seidel algorithm, as in,~\cite{GalkIvPZh06}  and Monte Carlo
analysis using the simulated annealing
technique.~\cite{MonteCarloBook} The energy minimization has been
performed as follows: we start with $\beta =0$, choose a simple
in-plane vortex as an initial condition, and numerically minimize
the energy then the value of $\beta $ is increased step by step.
This  method is working pretty fast and gives a good description of
the structure for continuous transitions, see below, while the Monte
Carlo method is important for the analysis of points where the
magnetic structure of the system is changing discontinuously, with a
coexistence of (metastable) states close to the transition point.

\begin{figure}
\includegraphics[width=\figwidth]{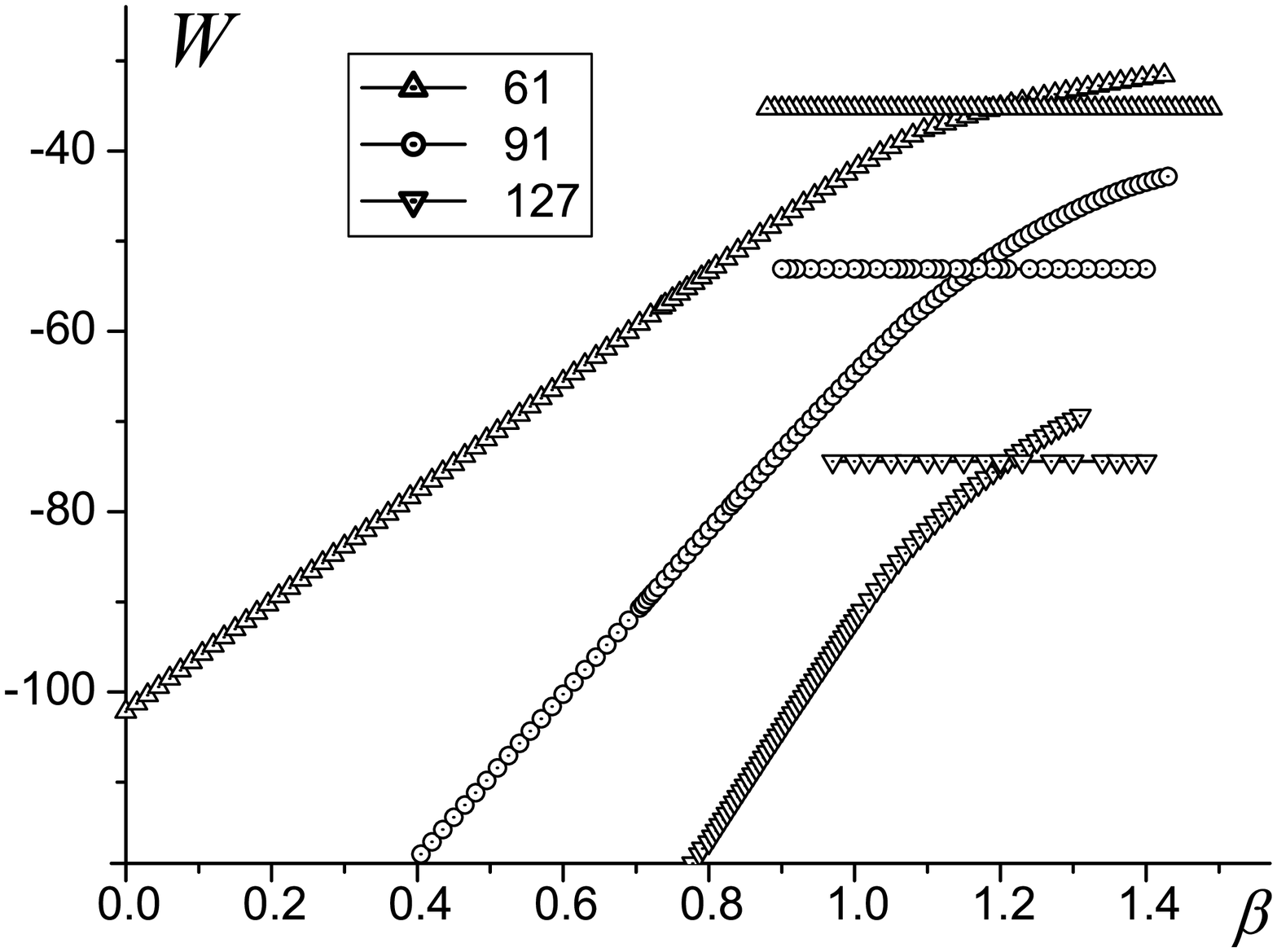}
\caption{\label{fig3}  The dependence of the energy  $W$ on the
anisotropy constant $\beta$ for clusters of different sizes (labels
in the legend denote the number of spins in the cluster), found by
numerical minimization of (\ref{eq1}). Horizontal lines denote the
energy of three-domain antiferromagnetic state, see Fig.~\ref{5}
while the sloped curves correspond to the energy of a vortex state.}
\end{figure}

Numerical calculations have been carried out for comparatively small
clusters shaped in the form of a regular hexagonal fragment of the
lattice consisting of 19, 37, 61, 91, and 127 nanoparticles. For all
studied systems, we have found an in-plane vortex in the ground
state at small enough $\beta$. The structure of the vortex changes
considerably  as $\beta$ increases and passes through two critical
values $\beta _1 $ and $\beta _2 $, see. Fig.~\ref{fig2} and the
detailed discussion below. Further, we have  found a prominent
transition at some value $\beta =\beta _t >\beta _{1,2} $ from the
vortex state to the state with a fragment of the antiferromagnetic
structure, and close to $\beta=\beta _t $ we have observed a
noticeable region of coexistence of vortex and antiferromagnetic
states, see Fig.~\ref{fig3}. The behavior of the energy as a
function of the anisotropy constant $\beta$ near this transition is
similar to that for a thermodynamic potential as a function of
temperature near first-order phase transition. On the other hand,
the dependence of the energy on $\beta$ did not exhibit visible
peculiarities at $\beta =\beta _{1,2} $, where a change of the
vortex structure has been detected.

\begin{figure}
\includegraphics[width=\figwidth]{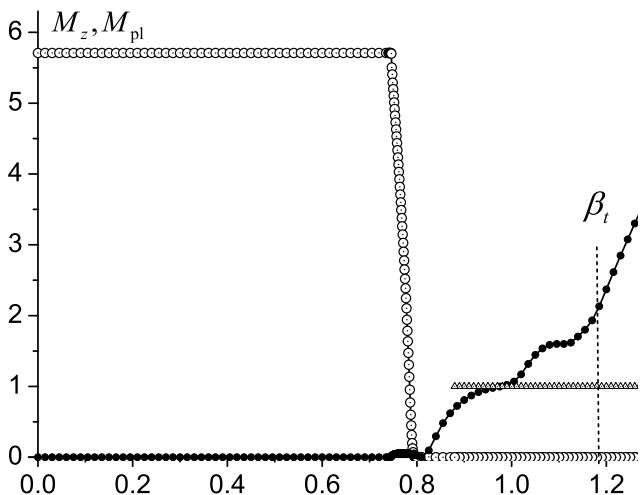}
\caption{\label{fig4} The in-plane and out-of-plane magnetic moment
  components  $M_z $ (full symbols) and $M_{\mathrm{pl}} $
(open symbols) vs the anisotropy constant $\beta $, in units of $\mu
_0 $, for a cluster with 61 particles. The details of behavior in
the region of vortex core reconstruction, $\beta_1 < \beta_2
<\beta$, here $\beta_1 \simeq  0.744$ $\beta_2  \simeq  0.786$, are
presented in Fig.~\ref{fig5}  on a different scale. Up triangles
label  the value $M_z = \mu_0$ for the antiferromagnetic state. }
\end{figure}

To analyze the structure and symmetry of the vortex core, after
completing the energy minimization for each value of $\beta$, we
have been calculating the value of the out-of plane component
$M_{z}$ of the total magnetic moment, as well as the length of the
planar component $M_{\mathrm{pl}} =\sqrt {M_x ^2+M_y ^2} $.  It
turns out that just these parameters are most sensitive to the
vortex structure and allow observing peculiarities of the vortex
core behavior, see Fig.~\ref{fig4}.

The behavior of the total magnetic moment of a particle array with a
vortex is rather complicated. At small anisotropy, the picture
remains the same as for isotropic particles,\cite{PolitiPiniTreug}
and a vortex with purely planar distribution of magnetic moments
is realized. In this case, accordingly, the total $z$-projection of the magnetic
moment vanishes, but the planar component of the total
magnetic moment $M_{\mathrm{pl}} $ is non-zero. With increasing
$\beta$, a nonzero value of $M_z$ emerges at some critical anisotropy $\beta =\beta _1 $. The fact that
$M_z \ne 0$ means that an out-of-plane core appears. However, with
the appearance of nonzero $M_z$, the planar component of the total
moment does not vanish immediately, i.e. within some finite interval
$\beta _1 <\beta <\beta _2 $ both in-plane and out-of-plane components of the moment are
nonzero, see Fig.~\ref{fig4}. The planar component vanishes for $\beta
>\beta _2 $, whereupon the vortex
state structure becomes more symmetric than that observed at $\beta
<\beta _2 $, see Fig.~\ref{fig2}. Such a symmetric vortex structure
is observed within a wide range of $\beta$, $\beta _2 <\beta < \beta
_t$ and the magnetic moment $M_z $ changes considerably while
$M_{\mathrm pl}$ remains zero. With the further increase of the
anisotropy, the vortex structure becomes an antiferromagnetic
structure similar to that which is found for strong perpendicular
anisotropy. The presence of boundaries  leads to the emergence of
three different domains of such structure, as dictated by the system
symmetry.

Note that the behavior of $M_z $ and $M_{\mathrm{pl}} $ is similar
to that of order parameters near second-order phase transitions.
This observation can be used to perform a symmetry analysis of the
transitions between the different vortex states. It is important to
obtain analytical results, due to the limited accuracy of numerical
data, but also because the numerical analysis is hindered near
transition points $\beta =\beta _{1,2} $ because of the ``critical
slowing down'' of relaxation similar to that found near a
second-order phase transition, which manifests itself in a
substantial increase in the numerical calculation time. Therefore,
we should study the possibilities of existence of such transitions
in our system from the viewpoint of symmetry.

\section{Symmetry analysis}

\begin{figure}
\includegraphics[width=\figwidth]{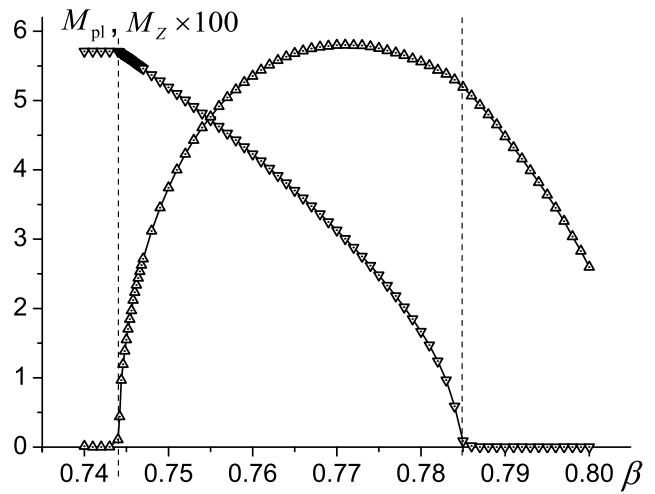}
\caption{\label{fig5} The in-plane and out-of-plane magnetic moment
  components  $M_z $ (up triangles) and $M_{\mathrm{pl}} $
(down triangles) vs the anisotropy constant $\beta $, in units of $\mu
_0 $, for a cluster with 61 particles, in the region of the vortex core
reconstruction. Note that the scales for $M_{\mathrm{pl}} $ and
$M_z $ differ by the factor of 100. }
\end{figure}

To describe the complex character of changing the vortex core
structure let us use symmetry arguments in line with the phase
transition theory of Landau. For both observed critical values of
anisotropy, the symmetry of state changes essentially. At $\beta \ge
\beta _1 $, when out-of-plane core emerges for the first time, the
sign of $M_z $ can be arbitrary, i.e., there is a spontaneous
breaking of $Z_{2}$ symmetry with respect to $M_{z}$ at
$\beta=\beta_{1}$. In contrast to that, the symmetry of the planar
distribution of magnetic moments does not change at this transition
point: within the range $\beta _1 <\beta <\beta _2 $ it remains the
same as for $\beta <\beta _1 $. At the other transition point
$\beta=\beta_{2}$ the situation is different: the dependence of the
out-of-plane component  $M_z (\beta )$ does not have any visible
peculiarities, while the planar component  $M_{\mathrm{pl}} $
vanishes at $\beta=\beta_{2}$ and remains zero for $\beta_{2}\le
\beta \le \beta _t $, i.e., up to the point of destruction of the
vortex state at $\beta =\beta _t$, see the detailed graph in
Fig.~\ref{fig5}.

\begin{figure}
\includegraphics[width=\figwidth]{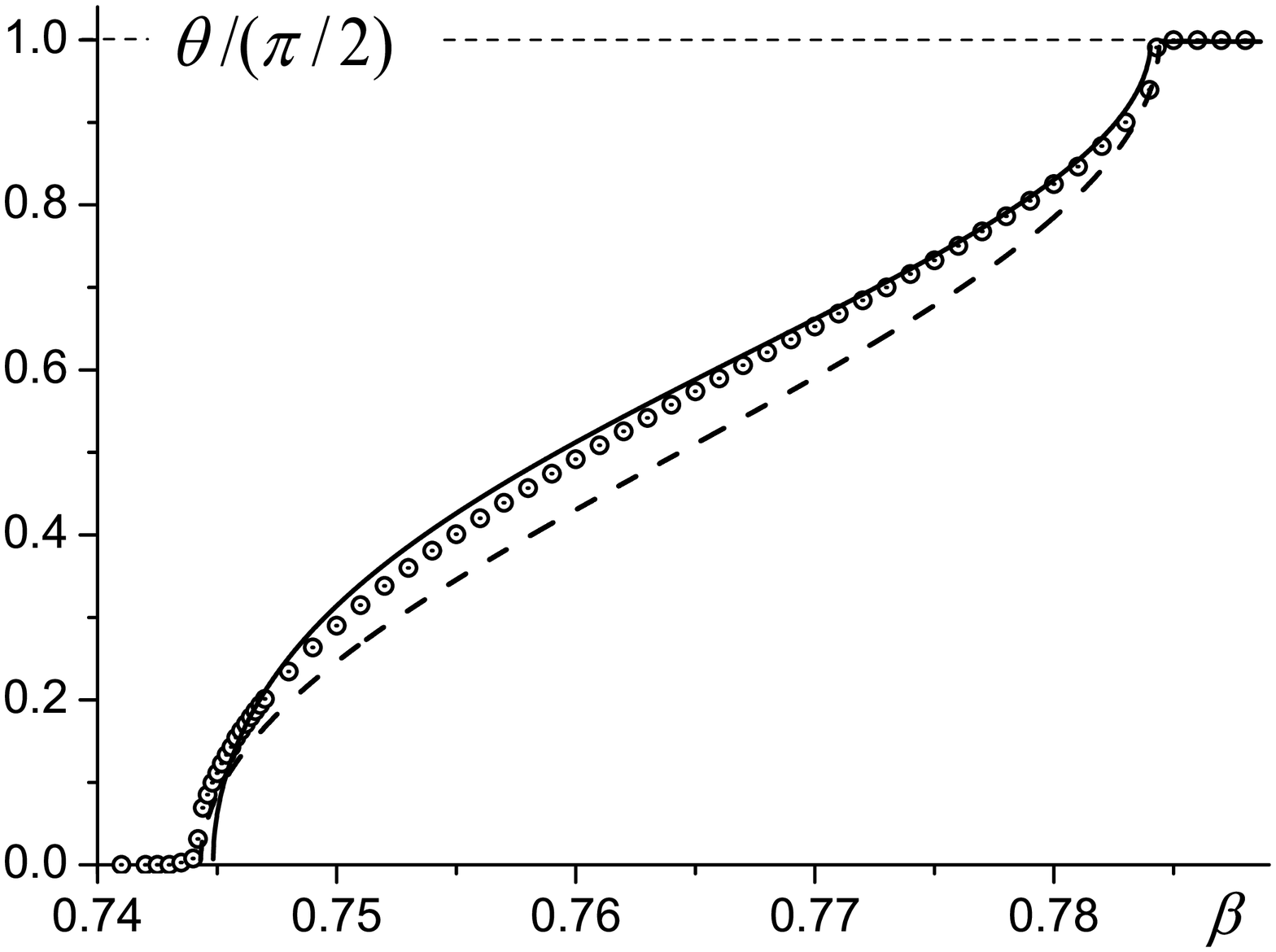}
\caption{\label{fig6}  The value of the angle $\theta _0 $
(normalized by $\pi /2)$ vs. the anisotropy constant $\beta$ for a
cluster with 61 particles in the region of the vortex core
reconstruction. Symbols present the results of numerical
calculations, while the dashed and solid lines show analytical
results from the phenomenological theory. The dashed line
corresponds to Eq.~(\ref{eq3}),  and the solid line shows the result
that follows from including corrections due to the sixth-order term
with $\overline{\beta}$, see the text. }
\end{figure}

For the ground state in the interval $\beta_{2}\le \beta \le \beta
_t$, we observe a higher symmetry of the moment distribution than
the vortex states with $M_{\mathrm{pl}} \ne 0$ at $\beta <\beta_2 $
(specifically, within the numerical accuracy of our simulations, we
observe the $C_{6}$ symmetry; it would be worth finding out whether
this symmetry is exact). Rather low symmetry of the vortex ground
state at $\beta <\beta _2 $ is caused by the presence of nonzero
$M_{\mathrm{pl}}$, which can be traced down to the presence of a
non-zero planar component of the central magnetic moment $\vec \mu_0
$. Obviously, the emergence of non-zero $M_z $ at $\beta>\beta_{1}$
is caused by $\vec \mu_0 $ coming out of plane. Writing down $\vec
\mu_0$ as $\vec{\mu}_{0} =\mu _0 \left( \sin \theta_0 \vec{e}_{z}
+\cos \theta_0 \vec{e}_{\mathrm{pl}}  \right)$, where
$\vec{e}_{\mathrm{pl}} $ lies in plane of the system, we obtain that
the transition at $\beta _1 $ is connected to the appearance of
non-zero $\theta _0$, with $\theta _0 =0$ at $\beta \le \beta _1 $
and $\theta _0 \ne 0$ at $\beta \ge \beta _1 $. As we pointed out,
the symmetry of the state with $M_z \ne 0$ is lower than for a
planar vortex, therefore the value of $\theta _0 $ serves as the
order parameter for the transition at $\beta =\beta _1 $. In this
case, one can expect that at $\beta \ge \beta _1$ the behavior of
the ``order parameter'' close to the transition is given by $ \theta
_0 \propto \sqrt {\beta -\beta _1 } $ and is characterized by
singular behavior, $d\theta _0 /d\beta \to \infty $ at $\beta \to
\beta _1 +0$. On the other hand, the hexagonal symmetry in the spin
distribution may appear only when the central moment is directed
strictly perpendicular to the system plane, i.e. at $\theta _0 =\pi
/2$. If at $\beta =\beta _2 $ the symmetry increases up to the
hexagonal one, the quantity $\vartheta _0 =\pi /2-\theta _0 $ should
serve as the order parameter for this transition.  We arrive at the
conclusion that the behavior of the out-of-plane component of the
central spin $\vec \mu_0 $, i.e., the dependence  $\theta _0 (\beta
)$, dictates the change of symmetry of the vortex state. The
information about the full $\theta_0 (\beta )$ dependence can be
obtained only numerically, but the presence of square-root
singularities at $\beta \to \beta _1 +0$ and $\beta \to \beta _2 -0$
is rather easily verified, see Fig.~\ref{fig6}.

The detailed analysis of the $\theta _0 (\beta )$ dependence allowes
one to present a closed phenomenological expression for the
``thermodynamic potential'' $\Phi$ that defines the behavior of
$\theta _0 (\beta )$ in a wide range of  $\beta $. Indeed, in line
with the Landau theory, this potential can be constructed in the
form of the expansion in powers of the order parameters, which are
$\theta _0 $ at $\beta \simeq \beta _1 $ or $\vartheta _0 =\pi
/2-\theta _0 $ at $\beta \simeq \beta _2 $. Equivalently, $\sin
\theta _0 $ and $\cos \theta _0 $ can be used instead of angles
$\theta _0 $ and $\vartheta _0 $. As odd degrees of $\mu _{0z} =\sin
\theta _0 $ are forbidden by the condition of the time reversal
invariance, and the simplest form of this energy is the following:
$\Phi =A\sin ^2\theta _0 +B\sin ^4\theta _0 $. It is easy to see
that, up to an inessential overall factor, the correct behavior is
provided by the expression
\begin{equation}
\label{eq2}
\Phi =\frac{1}{2}\left( {\beta _1 -\beta } \right)\sin ^2\theta _0
+\frac{1}{4}\left( {\beta _2 -\beta _1 } \right)\sin ^4\theta _0 ,
\end{equation}
which leads to the simple result:
\begin{equation}
\label{eq3}
\sin \theta _0 =\sqrt {\frac{\beta -\beta _1 }{\beta _2 -\beta _1 }} ,
\quad
\beta _1 \le \beta \le \beta _2 ,
\end{equation}
while $\theta _0 =0$ at $\beta
<\beta _1 $, and $\theta _0 =\pi /2$ at $\beta >\beta _2 $.

Such simple dependence  describes the numerical data fairly well,
see Fig.~\ref{fig6}. Deviation from the simple law given by
(\ref{eq3}) can be accounted for by adding the term $(\overline
\beta /6)\sin ^6\theta_0$ to the expansion (\ref{eq2}). As seen from
Fig.~\ref{fig6}, it provides a perfect description of the numerical
data at sufficiently small $\overline \beta $, typical values are
$\overline \beta \le 0.1(\beta _2 -\beta _1 )$.

\begin{figure}
\includegraphics[width=\figwidth]{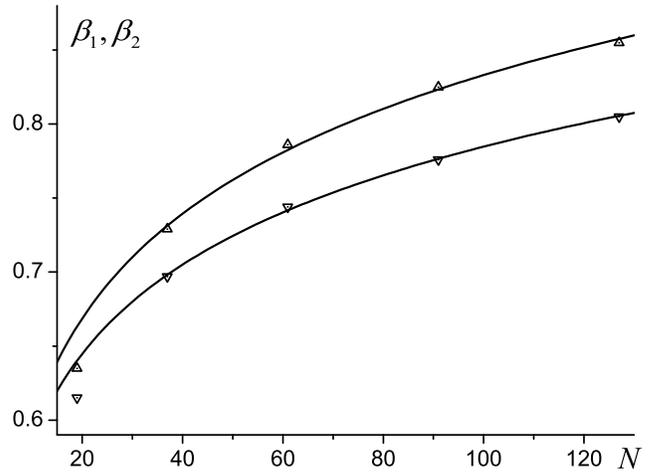}
\caption{\label{fig7}  The dependence of $\beta _2 $ (up triangles)
and $\beta _1 $ (down triangles) on the number of particles in the
cluster, lines present the dependence fitted by the logarithmic
function, see the text.}
\end{figure}

The critical values of the anisotropy constant $\beta _1 $ and
$\beta_2 $ grow with the increase of the cluster size $N$, see Fig.~\ref{fig7},
though for the studied values of $N$ this dependence is rather
slow. The numerical data for $N\ge 37$ can be well fitted by a
logarithmic dependence of the form $\beta _{1,2} =A_{1,2} +B_{1,2}
\ln N$, where $A_1 = 0.38397$, $B_1 = 0.08703$; $A_2 =0.36236$, $B_2
=0.10222$.

The value of $\theta _0 $ not only dictates the vortex core symmetry
but it also quantitatively defines the important vortex
characteristic, the total out-of-plane moment of the particle with
the vortex. Singularities in the $\theta _0 $ behavior at $\beta
=\beta _1$ are reflected in the $M_z \left( \beta \right)$
dependence, $M_z \propto \sin \theta _0 $ near this point. It is
worth noting that $M_{z}$ plays a special role in the dynamic
properties, namely, $M_z $ serves as a proper collective variable
describing the radial mode  (with the azimuthal number $m=0)$ of the
magnetization oscillations in the particle with a
vortex;~\cite{Galkin_m0Mode} the theory is in agreement with recent
experiment.~\cite{m0ModeExp2012}  Therefore, the presence of
singularities in $M_{z}(\beta)$ should manifest itself in the
behavior of an equivalent of this mode for the considered system.

In addition, the presence or absence of the vortex core is also
important for the properties of azimuthal modes with $m=\pm
1$.~\cite{Wysin96, Wysin+12} For a purely planar vortex, the modes
with $m=\pm 1$ form degenerate doublets, while the emergence of the
core leads to splitting of these doublets. Thus, one can expect  a
crucial impact of the vortex structure modification on the
properties of eigenmodes of the vortex-state particle, although a
detailed discussion of the dynamical properties of the particle with
the vortex is beyond the scope of this work.

\section{Summary and discussion}

To conclude, we have shown that high-symmetry hexagonal fragments of
a 2D closely-packed triangular lattice of magnetic particles contain
a vortex in the ground state, even for a small fragment size. The
vortex structure is very sensitive to the intrinsic anisotropy
$\beta $ of the particle. At small anisotropy, there is a purely
planar vortex. With the increase of $\beta$, the symmetry of the
vortex ground state lowers initially at some critical value $\beta
=\beta_1$, and then increases to a high sixfold axial symmetry above
another critical value $\beta = \beta _2 >\beta _1 $. It is worth to
note that those two transformations bear a similarity to
second-order phase transitions. Both transitions take place at
sufficiently weak anisotropy, the dimensionless parameters
$\beta_{1,2} $ do not exceed one. This value is essentially smaller
than the easy-plane anisotropy of a planar array induced by the
demagnetization field with the characteristic value $\beta
_{\mathrm{array}} \sim 10$, see Ref.~\onlinecite{Bondarenko+prb12}.
Actually, this anisotropy is smaller than it is necessary to create
the perpendicular magnetization of a cylindrical magnetic dot.

An important challenge in the physics of magnetic vortices is to
find ultra-small (smaller than $100$~nm) systems with vortices in
the ground state, this problem is of great interest for both
fundamental physics and applications. In addition to lithographic
magnetic materials, where the particle size is of the order of tens
of nanometers,\cite{chou94,Meier98,Ross} the proposed theory is
applicable to other 2D systems with anisotropic particles having
magnetic or electrical dipole moment.~\cite{dipObzorUFN} One could
expect that if an array can be composed from small enough particles,
having finite magnetic or electric dipole moment, the vortex state
will be present for arrays 10-20 times larger than the particle
size. Such systems can be realized for composite magnetic materials,
for example, for granular magnets with the content of the magnetic
component less than the percolation threshold, where the exchange
interaction between nanometer-sized grains is anomalously small.
Another example is the inhomogeneous state arising in the vicinity
of the metal--insulator transition in doped manganites, which
involves small particles of the ferromagnetic (metallic) phase
distributed over a nonmagnetic host; their physical properties are
determined to a large extent by the dipolar interactions between
these particles.~\cite{Krivoruchko} The experimental implementation
of the artificial crystals, in which particles with magnetic moments
of the order of $10^{3}$ Bohr magnetons form an ordered lattices,
has been reported recently.~\cite{NanopartCrysPRL10} As one more
example, it is instructive to mention a new class of materials,
namely, molecular crystals formed by  high-spin molecules. The total
magnetic moment of such a molecule can be as high as dozens of Bohr
magnetons, but the exchange interaction between magnetic moments of
different molecules is almost negligible.~\cite{Wernsdorfer01} Note
also so-called dense phases formed by nanometer-sized magnetic
particles moving freely in a liquid (that is the standard situation
for a ferrofluids).~\cite{ferrofluid} For all these systems with a
particle size of the order of nanometers the vortices described here
can be present for objects as small as dozen of nanometers; those
are, to the best of our knowledge, the smallest vortex-bearing
systems discussed in the literature.

It is worth noting that the presence of a vortex ground state for
such small systems and the transitions with the vortex core
reconstruction is a consequence of the high (hexagonal) symmetry of
the array. For square or rectangular arrays the vortex state appears
for large enough arrays only.~\cite{PolitiPiniTreug} As we found,
for an array shaped as a regular triangle, the vortex state could be
present for small arrays, but with the increase of the anisotropy the
vortex remain coreless all the way till the transition to antiferromagnetic
state. The two-dimensional nature of an array is also quite
important. Thus, two-dimensional closely-packed arrays of magnetic
particles represent  vortex-bearing systems with potentially
small sizes and offer a unique possibility for manipulating the
symmetry and structure of the vortex core.

We are thankful to V.~G. Baryakhtar, A.~K. Kolezhuk and V.~F.
Kovalenko for useful discussions. This work was partly supported by
the Government of Ukraine, State Program ``Nanotechnologies and
Nanomaterials'' project no. 1.1.3.27.

\end{document}